# Coherently induced quantum correlation in a delayed-choice scheme

Byoung S. Ham
School of Electrical Engineering and Computer Science, Gwangju Institute of Science and Technology
123 Chumdangwagi-ro, Buk-gu, Gwangju 61005, South Korea
(Submitted on July 30, 2023; bham@gist.ac.kr)

**Abstract**
Quantum entanglement is known as a unique quantum feature that cannot be obtained by classical physics. Over the last several decades, however, such an understanding on quantum entanglement might have confined us in a limited world of weird quantum mechanics. Unlike a single photon, a definite phase relation between paired photons is the key to understanding quantum features. Recently, an intuitive approach to the otherwise mysterious quantum features has emerged and shined a light on coherence manipulations of product-basis superposition via selective measurements. Here, a coherence manipulation is presented to excite polarization-path correlation using Poisson-distributed coherent photons for a classically excited joint-phase relation of independent local parameters. For this, linear optics is used for the preparation of the polarization-basis randomness, and a gated heterodyne detection technique is adopted for the selective measurement of polarization bases. As a result, the nonlocal quantum feature is now coherently understood in a deterministic way.

Quantum superposition is the heart of quantum mechanics whose mysterious quantum feature is in the wave-particle duality based on probability amplitudes of a single photon or particle [1,2]. In the wave-particle duality or complementarity theory of quantum mechanics, a quantum measurement plays an essential role to determine a photon's nature [3]. Wheeler's delayed-choice thought experiments are about measurement-dependent photon characteristics in an interferometric system [4]. In the quantum eraser [4], the cause-effect relation of a single photon or a pair of entangled photons has been violated in the delayed-choice scheme [5]. Thus far, the delayed-choice quantum eraser has been conducted using various photon natures of thermal [6], coherent [7-9], and entangled photons [5,10-13]. Recently, coherence solutions of the quantum eraser have been derived from both entangled [14] and coherent photons [15] to understand the fundamental physics of measurement-based quantum features. Like nonlocal correlation [16-21], the violation of the cause-effect relation in the quantum eraser originates in measurement-event modifications in a delayed-choice scheme, resulting in the second-order quantum superposition between product bases at the cost of a 50 % photon loss [14,15]. Without measurement-event modifications, no nonlocal quantum feature cannot be excited or observed. Here, a coherently induced quantum correlation is presented for the delayed-choice quantum eraser using attenuated light to understand the fundamental physics of nonlocal quantum features. For this, linear optics is used for the preparation of polarization-basis randomness, and a gated heterodyne detection technique is adopted for the selection of a certain group of polarization-product bases, resulting in the joint-parameter relation of the quantum feature.

For the EPR paradox [16], nonlinear optics-based nonclassical lights have been used for nonlocal correlation between space-like separated paired particles, violating local realism [17-20]. Thus, quantum entanglement has been understood as a mysterious quantum feature that cannot be accomplished by any classical means. Unlike the delayed-choice quantum eraser relating to a single particle [3], it is commonly accepted that quantum entanglement is a probabilistic process by such as spontaneous parametric down-conversion (SPDC) [22,23] or a controlled π-phase shift [24]. Thus, coherent photons have been generally excluded from the potential candidates of the nonlocal quantum features. Recently, such conventional beliefs on the nonlocal correlation have been seriously challenged by a new interpretation, where coherent photons can be directly used to excite nonlocal quantum features [25]. Such a coherence approach is based on coincidence detection-caused measurement modifications, resulting in an inseparable intensity product [14,26]. For the measurement-based coherence approach using an attenuated laser, a Franson-type correlation has been successfully analyzed for the ensemble decoherence among long-life individual single photons in an asymmetric Mach-Zehnder interferometer (MZI) [25,26]. Regarding Poisson-distributed photon statistics of an attenuated laser, only doubly-bunched photons can be conditionally taken for the coincidence measurements with a statistically negligible error [27].

Figure 1(a) shows the schematic of the coherently induced quantum correlation using Poisson-distributed coherent photon pairs from a continuous wave (cw) laser. For random distribution of orthogonally polarized photons inside the MZI, a 22.5°-rotated half-wave plate (HWP) is placed before the MZI. In both MZI paths, each photon pair is frequency-modulated at an opposite ($\pm \delta f$) detuning mode, resulting in a frequency-path correlated photon pair, as shown in Fig. 1(b). For the $j^{th}$ frequency pair at $f_\pm = f_0 \pm \delta f_j$, a pair of synchronized acousto-optic modulators (AOMs) is used. To satisfy the random detuning pair at $f_\pm$, the modulation rate of AOMs should be higher than the coincidence detection rate. Like SPDC-generated photon pairs [22,23], the coherently excited frequency-path correlation satisfies the same polarization-path correlation relation by the PBS. Due to the BS randomness, the polarization basis in each MZI output port is also random (see the top chart of Table 1). For the measurement modification of the output photons, a gated heterodyne detection technique is adopted for the coincidence measurements. By the heterodyne detection, the homodyne case of the bunched photons in either MZI path is conditionally removed. As a result, all generated coherent photon pairs used for measurements in Fig. 1 satisfy the polarization-path correlation at an equal ratio. These polarization-correlated photon pairs are for the same polarization basis at an opposite frequency, satisfying the entanglement relation: $|H^+\rangle_A|H^-\rangle_B + |H^-\rangle_A|H^+\rangle_B$; $|V^+\rangle_A|V^-\rangle_B + |V^-\rangle_A|V^+\rangle_B$; $|H^\pm\rangle_A|H^\mp\rangle_B + |V^\pm\rangle_A|V^\mp\rangle_B$. Here, superscripts '+' and '−' indicate $f_+$ and $f_-$, respectively. The selection efficiency by the coincidence measurements is thus 25 % of generated photon pairs. (see the bottom chart of Table 1). Understanding the role of selective measurements is the key to understanding nonlocal correlation [13,14,21,25,26]. By the PBS, no φ-dependent fringe is possible for both output intensities $I_A$ and $I_B$: basis randomness [28,29].

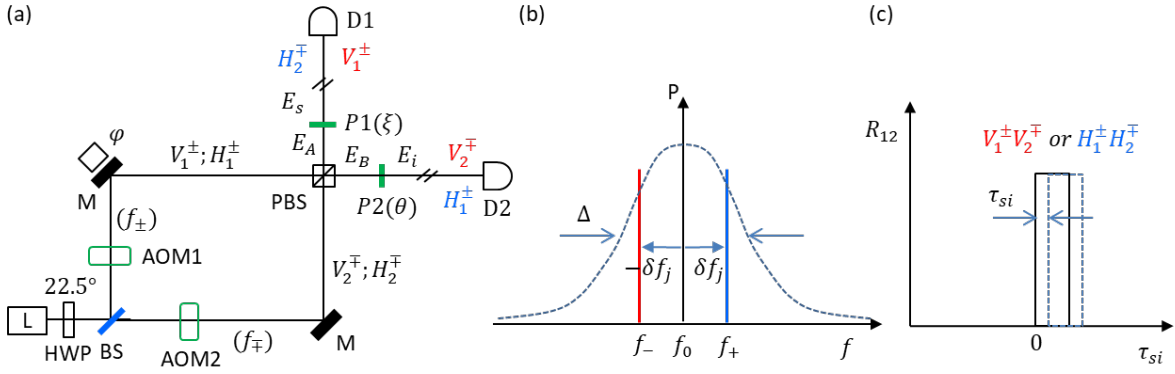

**Fig. 1. Coherently manipulated nonlocal quantum scheme using an attenuated laser.** (a) Schematic of coherence manipulations for quantum correlation using heterodyne detection-based coincidence measurements. (b) A coherent photon pair generated from Fig. 1(a). (c) Heterodyne detection resulting intensity product in (a). BS: balanced beam splitter, AOM: acousto-optic modulator, BS: beam splitter, D: photodetector, M: mirror P: polarizer, PBS: polarizing BS. $V_j^\pm$ ($H_j$): vertically (horizontally) polarized photon in path j, where the sign $+(-)$ indicates positively (negatively) detuned frequency from $f_0$. $\tau_{si}$ is the time delay between D1- and D2-generated electrical pulses in a coincidence counting module. Δ is the spectral bandwidth of the AOM-generated diffraction lights.

**Table 1. Polarization-basis correlations for Fig. 1.** The letter 'C' ('X') indicates heterodyne detection-based correlation (exclusion).

|  |  | MZI path 1 | |
|---|---|---|---|
|  |  | $H_1$ | $V_1$ |
| MZI path 2 | $H_2$ | C | C |
|  | $V_2$ | C | C |

|  |  | Output path A | |
|---|---|---|---|
|  |  | $H_1$ | $V_2$ |
| Output path B | $H_2$ | C | X |
|  | $V_1$ | X | C |

For the second-order intensity correlation between two space-like separated output photons in Fig. 1, conventional coincidence measurements are conducted for gated heterodyne detection. Here, the gated mode of the heterodyne signals is to avoid the time averaging of signals. For this, the resolving time of the photon detector must be shorter than the gate window.

For local measurements in both MZI output ports in Fig. 1(a), the particle nature of a photon determined by PBS is erased by the inserted polarizer P1 (ξ) or P2 (θ) in a delayed-choice manner [5,8,10]. In the quantum eraser limited to the first-order intensity correlation, there is no difference between a single photon [9,15] and continuous wave (cw), as observed [15,30]. This is the same mechanism as the self-interference of a single photon in a double-slit system [1,27,31]. As analyzed in ref. [15], the function of Ps is to convert the orthogonal polarization bases into a common basis, resulting in the retrieval of the indistinguishable photon characteristics of the wave nature [8,13,21]. In this P-based measurement process, another cost of photon loss at 50 % is inevitable. Thus, the violation of the cause-effect relation is due to the measurement-event modification [15]. This kind of measurement modification has also been discussed for the Franson-type nonlocal correlation [25,26] as an essential feature of quantum mechanics distinguished from classical physics.

The gated heterodyne detection adopted for the joint measurement between two local detectors D1 and D2 in Fig. 1 is to induce polarization (frequency)-path correlation via freezing (coincident) a time window of the measured photons. This product-basis selection is the first step toward the manipulation of the quantum feature in the present paper. For the MZI output ports in Fig, 1(a), a frequency pair randomly generated by AOMs is selectively measured for the frequency (polarization) basis, resulting in the entanglement relation, as shown above. In the quantum eraser, the goal of coherence manipulations of the product-basis sets is for randomness-based quantum superposition. Due to this basis-randomness, thus, local intensities become uniform.

In Fig. 1(a), the amplitudes of the PBS-resulting distinguishable photon pairs are denoted by $E_A$ and $E_B$, whose resulting polarization (frequency)-path correlations are color-matched in red or blue for the gated heterodyne detection. As mentioned above, all cross-color-matched pairs should be excluded from measurements for heterodyne detection. Thus, the coincidence measurement by the gated heterodyne detection technique plays an essential role in selective measurements. By the PBS of the MZI [28,29], no fringe results in both local intensities $I_A$ and $I_B$:

$$\langle I_A^j \rangle = \frac{\langle I_0 \rangle}{2} \langle (-V_1^\pm e^{\pm i 2\delta f_j \tau} + H_2^\mp)(-V_1^\mp e^{\mp i 2\delta f_j \tau} + H_2^\pm) \rangle = \langle I_0 \rangle, \tag{1}$$

where $I_k = E_k E_k^*$ and $E_k$ is the amplitude of the $j^{th}$ single photon. The MZI path-length control affects differently for all $\delta f_j$-detuned photons: $\varphi_j = \pm \delta f_j \tau$, where $\tau$ is the path-length difference-caused time delay between the paired photons. Likewise,

$$\langle I_B^j \rangle = \frac{\langle I_0 \rangle}{2} \langle (H_1^\pm e^{\pm i 2\delta f_j \tau} + V_2^\mp)(H_1^\mp e^{\mp i 2\delta f_j \tau} + V_2^\pm) \rangle = \langle I_0 \rangle. \tag{2}$$

Equations (1) and (2) show the random basis-caused particle nature of photon characteristics, resulting in a uniform local intensity.

By the inserted polarizers Ps (ξ; θ) in both MZI output ports, however, the output fields $E_A$ and $E_B$ are modified to the polarization projection onto a common basis determined by P's rotation angles ξ and θ, respectively. Thus, the final amplitudes of both MZI output photons are represented as:

$$E_s^j = \frac{E_0}{2}(-V_1^\pm \sin\xi\, e^{\pm i 2\delta f_j \tau} + H_2^\mp \cos\xi), \tag{3}$$

$$E_i^j = \frac{iE_0}{2}(H_1^\pm \cos\theta\, e^{\pm i 2\delta f_j \tau} + V_2^\mp \sin\theta), \tag{4}$$

where the polarization bases $V_j^\pm$ and $H_j^\pm$ are just to indicate their polarization origins. The corresponding intensities of Eqs. (3) and (4) show the quantum eraser:

$$\langle I_s \rangle = \frac{\langle I_0 \rangle}{2} \langle (-V_1^\pm \sin\xi\, e^{\pm i 2\delta f_j \tau} + H_2^\mp \cos\xi)(-V_1^\mp \sin\xi\, e^{\mp i 2\delta f_j \tau} + H_2^\pm \cos\xi) \rangle$$

$$= \frac{\langle I_0 \rangle}{2} \langle 1 - \sin 2\xi \cos(2\delta f_j \tau) \rangle, \tag{5}$$

$$\langle I_i \rangle = \frac{\langle I_0 \rangle}{2} \langle \left( H_1^\pm \cos\theta e^{\pm i2\delta f_j \tau} + V_2^\mp \sin\theta \right)\left( H_1^\mp \cos\theta e^{\mp i2\delta f_j \tau} + V_2^\pm \sin\theta \right) \rangle$$
$$= \frac{\langle I_0 \rangle}{2} \langle 1 + \sin 2\theta \cos(2\delta f_j \tau) \rangle. \qquad (6)$$

Near coincidence detection at $\tau \sim 0$, both local intensities show the polarizer-dependent fringes. For the violation of cause-effect relation, the MZI path lengths are simply set to be beyond the light cone. For the nonlocal condition, the distances to both parties from the MZI are also set to be beyond the light cone. If $\tau \gg \Delta^{-1}$, the mean intensities become $\langle I_s \rangle = \langle I_i \rangle = \frac{\langle I_0 \rangle}{2}$, i.e., the classical lower bound, regardless of $\xi$ and $\theta$. This is a pure coherence ensemble effect by AOM-induced coherent photon pairs to the MZI. Although the MZI is tolerant to the laser bandwidth $\delta_{laser}$, the condition $\delta_{laser} \ll \Delta$ should be satisfied to have the frequency-path correlation.

Now, we solve Eqs. (3) and (4) for the joint-intensity correlation via a gated detection of the heterodyne signals for coincidence measurements. The coincidence detection-caused intensity product between two local detectors D1 and D2 in Fig. 1 is as follows:

$$\langle R_{si}(\tau_{si}) \rangle = \langle (E_s E_i)(E_s E_i)^*(cc) \rangle$$
$$= \langle \frac{I_0^2}{4} e^{-2\tau_{si}/\tau_c} \left( -V_1^\pm \sin\xi e^{\pm i2\delta f_j \tau} + H_2^\mp \cos\xi \right)\left( H_1^\pm \cos\theta e^{\pm i2\delta f_j \tau} + V_2^\mp \sin\theta \right)(cc) \rangle$$
$$= \frac{\langle I_0^2 \rangle}{4} e^{-2\tau_{si}/\tau} \langle \cos^2(\xi + \theta) \rangle, \qquad (7)$$

where $\tau_c$ is the ensemble coherence equivalent to $\Delta^{-1}$. In Eq. (7), the gated heterodyne detection plays a critical role in the coherently excited nonlocal fringe for the joint-phase relation of local polarizers $\xi$ and $\theta$, where the frequency-detuning effect is perfectly and automatically removed [32]. Thus, the coherence solution of the nonlocal correlation for Fig. 1 is successfully derived for the gated heterodyne detection-based coincidence measurements using a coherently manipulated cw laser. Surprisingly, this nonlocal correlation is nothing but due to measurement modification-caused product-basis superposition for the random polarization bases. This is of course inherently satisfied for SPDC-generated entangled photons [25,32]. Thus, the Poisson-distributed coherent photons are successfully manipulated for the nonlocal quantum feature of inseparable intensity products, as in conventionally observed using entangled photons [12,13,18-20]. Here, it should be noted that a definite phase relation between paired photons is the prerequisite of the product-basis superposition for the nonlocal correlation between the AOM-induced frequency-correlated photons [14,15].

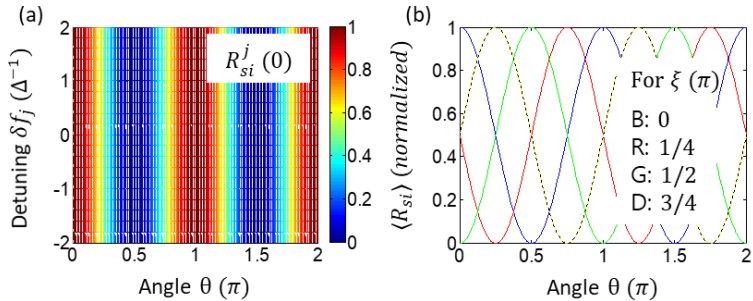

Fig. 2. Numerical calculations for $R_{si}(\tau_{si})$ using Eqs. (3) and (4). (a) Detuning independent $R_{si}^j(0)$. (b) Joint phase relation of $\langle R_{si}(0) \rangle$.

Figure 2 shows numerical calculations for the analytical solution of Eq. (7). To understand the detuning independent two-photon correlation in Eq. (7), Eqs. (3) and (4) are directly used for calculations, where the AOM-induced photon bandwidth $\Delta$ is set to be 1 MHz. For the numerical calculations, the frequency range of each photon pair is set for $-2\Delta \leq \delta f_j \leq 2\Delta$. The frequency step is set at $\Delta/5$. Figure 2(a) shows the detuning $\delta f_j$-independent individual $R_{si}^j(0)$. Figure 2(b) is the average of Fig. 2(a) for different $\xi$s, satisfying Eq. (7). Thus, the coherently excited nonlocal quantum feature of Eq. (7) is numerically confirmed, where it is ideal with a narrow bandwidth laser L. The AOM frequency scanning is not necessary due to the $\delta f_j$ independency, as shown in Fig. 2(a). Setting the symmetric AOMs is just to mimic the SPDC case in Fig. 1(b) for a helpful understanding of the coherence approach. Due to the polarization correlation between measured photon pairs in both parties, resulting in the color-matched photon pairs, even a cw case might have no difference from the single-photon case if measurement modifications were performed for the inseparable product basis.

In conclusion, a coherence version of the nonlocal correlation was theoretically investigated in a delayed-choice quantum eraser scheme using an attenuated cw laser. For the coherence manipulation of the frequency (polarization)-path correlated photon pairs, a pair of synchronized AOMs was used. A gated heterodyne detection technique was adopted for coincidence measurements to selectively choose a certain group of polarization-product bases. The *a priori* particle nature of a photon from the MZI was erased by the delayed-choice quantum eraser at the cost of 50 % photon loss. The role of the quantum eraser was to induce quantum superposition between particular product bases via gated heterodyne detection. As a result, a particular polarization-product basis at an opposite frequency detuning mode was selected, resulting in the detuning-independent joint phase relation of the nonlocal quantum feature. If coincidence was not satisfied, the nonlocal correlation deteriorated due to the ensemble decoherence. However, with a narrow bandwidth or frequency-fixed AOMs, the coincidence has no practical meaning, resulting in deterministic processing. From the coherence solution of the nonlocal fringe, the quantum feature was coherently understood for the selective measurement process, resulting in the product-basis superposition. Compared to the first-order intensity fringes of the single photon's self-interference, the coherently derived nonlocal intensity-product fringes between space-like separated parties represented the second-order quantum superposition between paired photons. In this coherence interpretation, the frequency (polarization) correlation between paired photons is deterministic even though the frequency (polarization) basis is random for individual pairs.

**Funding:** This research was supported by the MSIT (Ministry of Science and ICT), Korea, under the ITRC (Information Technology Research Center) support program (IITP 2023--2022-2021-0-01810) supervised by the IITP (Institute for Information & Communications Technology Planning & Evaluation). BSH also acknowledges that this work was also supported by GIST GRI-2023.

**Author contribution:** BSH solely wrote the paper.

**Competing Interests:** The author declares no competing interest.

**Data Availability Statement:** No Data associated in the manuscript.